\documentclass[aps,showpacs,twocolumn,dvips]{revtex4}
\usepackage{feynmp}
\usepackage{amssymb}
\usepackage{epsfig}
\usepackage{graphicx}
\usepackage{subfigure}
\usepackage{mathrsfs}
\usepackage{hyperref}
\usepackage{cleveref}
\usepackage{cases}
\usepackage{bbm}
\usepackage{xcolor}
\usepackage{tabularx}
\usepackage{array}
\usepackage{multirow}
\usepackage{tikz}
\usepackage{calc}
\usepackage{pifont}%

\usepackage{lmodern}
\usepackage{textcomp}
\usepackage{makecell}
\usepackage{threeparttable}

\begin{document}


\title{Topological superconducting transition driven by time-reversal-symmetry breaking}

\date{\today}

\author{Jing Wang}
\altaffiliation{E-mail address: jing$\textunderscore$wang@tju.edu.cn}
\affiliation{Department of Physics, Tianjin University, Tianjin
300072, P. R. China}

\begin{abstract}
Three-dimensional line-nodal superconductors exhibit nontrivial
topology, which is protected by the time-reversal symmetry. Here we
investigate four types of short-range interaction between the
gapless line-nodal fermionic quasiparticles by carrying
renormalization group analysis. We find that such interactions can
induce the dynamical breaking of time-reversal symmetry, which
alters the topology and might lead to six possible distinct
superconducting states, distinguished by the group representations.
After computing the susceptibilities for all the possible
phase-transition instabilities, we establish that the
superconducting pairing characterized by $id_{xz}$-wave gap symmetry
is the leading instability in noncentrosymmetric superconductors.
Appropriate extension of this approach is promising to pick out the
most favorable superconducting pairing during similar topology-changing transition
in the polar phase of $^3$He.
\end{abstract}


\maketitle


\section{Introduction}

Superconductivity has been studied for more than one century.
Ordinary phonon-mediated superconductors have an isotropic $s$-wave
superconducting (SC) gap, which suppresses fermionic excitations at
low energies \cite{BCS1957PR,Tinkham1996Book}. Cuprate
superconductors possess an anisotropic $d_{x^2-y^2}$-wave SC gap
\cite{Lee2006RMP} that hosts four nodal points. In some cases, pure
$d_{x^2-y^2}$-wave SC state may undergo a quantum phase transition
(QPT), entering into another SC state which has a distinct gap
symmetry \cite{Vojta2000PRL}. The quantum criticality of such a QPT
is governed by the interaction between nodal quasiparticles and
quantum fluctuation of new SC order parameter,
and the proximity to $d_{x^2-y^2}+id_{xy}$-wave SC phase was argued
\cite{Vojta2000PRL} to account for the observed marginal Fermi
liquid behavior in optimally doped
Bi$_2$Sr$_2$CaCu$_2$O$_{8+\delta}$ \cite{Valla1999Science}.

Experiments reveal that, some three-dimensional (3D) heavy fermion
compounds~\cite{Bonalde2005PRL,Izawa2005PRL,Gasparini2010JLTP} and 
iron-based superconductors~\cite{Reid2010PRB,Reid2010PRL} exhibit a line-shape
nodal structure in the SC gap. Examples are 
$\mathrm{CePt_3Si}$~\cite{Bonalde2005PRL,Izawa2005PRL}, UCoGe~\cite{Gasparini2010JLTP}, 
and $\mathrm{Ba(Fe_{1-x}Co_x)_2As_2}$~\cite{Reid2010PRB,Reid2010PRL}. 
Due to the presence of gap vanishing lines, there are fermionic excitations
even at ultra-low energies~\cite{Sigrist1991RMP,Matsuda2006JPCM}.
Different from 2D point-nodal superconductors, 3D line-nodal
superconductor may host striking topological SC state~\cite{Matsuura2013NJP}. 
Of particular interest is the nodal
centrosymmetric superconductor in which the absence of inversion
symmetry allows for the mixing of even and odd parities.

\begin{figure}
\centering
\includegraphics[width=3.0in]{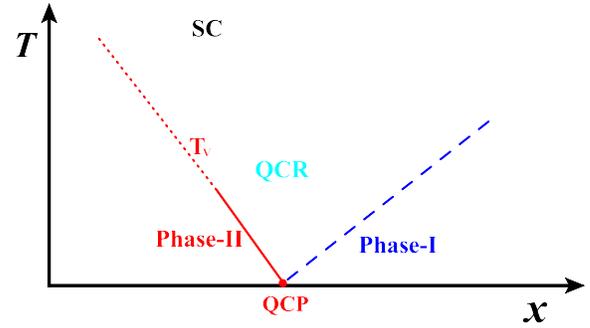}
\vspace{-0.13cm}\caption{(Color online) Schematic phase diagram on
$x$-$T$ plane, where $x$ is non-thermal tuning parameter and $T$
temperature. A putative QCP separates two distinct superconducting
phases, i.e., Phases I and II. $\mathcal{T}$-symmetry is respected
in Phase I and broken in Phase II.}\label{Fig_phase}
\end{figure}

The stability of the topological SC state in nodal
noncentrosymmetric superconductors is protected by the time-reversal
($\mathcal{T}$) symmetry \cite{Matsuura2013NJP}. This topology will
be changed once $\mathcal{T}$-symmetry is broken \cite{Moon2017PRB}.
$\mathcal{T}$-symmetry breaking ($\mathcal{T}$SB) drives the system
into a distinct SC phase. This transition is characterized by the
simultaneous changes of both symmetry and topology,
at which the conventional symmetry-breaking is a crucial ingredient
in reconstructing the band structure and leading to topological changes~\cite{Fradkin2009PRL,Vafek2014PRB,Wang2017QBCP}.
A schematic phase diagram is presented in Fig.~\ref{Fig_phase}. Two distinct SC
phases, i.e., Phase I and Phase II, are separated by a quantum
critical point (QCP) in the $x$-axis. Phase I is a
$\mathcal{T}$-symmetry protected topological SC phase, whereas Phase
II a $\mathcal{T}$SB SC phase. The gap structure of $\mathcal{T}$SB
SC state is not unique, and should be specified by the
representations of the underlying group of the system. For instance,
for a system described by $C_{4v} \times \mathcal{T} \times
\mathcal{P}$ group, where $C_{4v}$ is the tetragonal symmetry
(inversion-symmetry broken) and $\mathcal{P}$ is particle-hole
symmetry, there are six possible $\mathcal{T}$SB SC states
\cite{Moon2017PRB}, distinguished by the gap structure. The order
parameters for these six states are summarized in
Table~\ref{table_source}. As it concerns the question on whether the phases shown in Fig.~\ref{Fig_phase}
are topological or not, that crucially depends upon the corresponding order parameters
and the specific broken symmetries. We therefore are allowed to extract
the non-trivial changes of topological phases from investigating the related
traditional symmetry breakings~\cite{Fradkin2009PRL,Vafek2014PRB,Wang2017QBCP,Moon2017PRB}.

\renewcommand\arraystretch{2.0}
\begin{table*}[htbp]
\caption{There are six different $\mathcal{T}$SB SC phases, which
have different gap structures and different coupling matrices for
the coupling between fermions and source terms. These
$\mathcal{T}$SB SC phases with inversion-symmetry breaking are
associated with distinct sorts of continuum representations
(coupling matrices) near their nodal lines owing to the qualitative
difference of nodal structures \cite{Moon2017PRB}. There are six
possible states: 1) SC-I ($is$-pairing); 2) SC-II ($ig$-pairing); 3)
SC-III ($id_{x^2-y^2}$-pairing); 4) SC-IV ($id_{xy}$-pairing); 5)
SC-V ($id_{xz}$-pairing); and 6) SC-VI ($id_{yz}$-pairing).}
\label{table_source} \vspace{0.3cm} 
\begin{tabular}{p{3.5cm}<{\centering} p{2.1cm}<{\centering} p{2.35cm}<{\centering}
p{2.0cm}<{\centering} p{2.35cm}<{\centering} p{2.0cm}<{\centering} p{2.0cm}<{\centering}}
\hline \hline Order parameter ($\Delta_i$)  & SC-I & SC-II & SC-III
& SC-IV
& SC-V & SC-VI \\
\hline Coupling matrices ($\mathcal{M}_i$) & $\tau^y$ &
$\frac{1}{4}\sin(4\theta_{\mathbf{k}})\tau^y$ &
$\cos(2\theta_{\mathbf{k}})\tau^y $ &
$\frac{1}{2}\sin(2\theta_{\mathbf{k}})\tau^y $ &
$\cos(\theta_{\mathbf{k}})\tau^y $
& $\sin(\theta_{\mathbf{k}})\tau^y$ \\
\hline \hline
\end{tabular}
\end{table*}

Two interesting questions arise naturally. What is the driving force
of $\mathcal{T}$SB? How could one determine the most favorable
$\mathcal{T}$SB SC phase among the six candidates?

To answer these questions, in this paper we will consider four
types of short-range interactions between the gapless fermionic
quasiparticles excited near the SC gap nodal lines. After performing
one-loop renormalization group (RG) analysis \cite{Wilson1975RMP,
Polchinski9210046,Shankar1994RMP}, we derive the coupled flow
equations of all the model parameters, including the fermion
velocities and the coupling constants. Based on the RG results, we
show that all the four coupling constants diverge at low energies.
This indicates that the system becomes unstable and should enter
into a distinct new phase. This new phase could be one of the six
$\mathcal{T}$SB SC phases.

It should be noted that RG flows of coupling parameters do not
provide sufficient information to determine the leading ordering
tendency~\cite{Chubukov2016PRX}. To find out the dominant
instability, we will study all the possible ordered phases
(restricted to the charge channel), and evaluate the corresponding
susceptibilities for each ordered phase based on the RG flow
equations of model parameters. Our analytical and numerical
calculations indicate that, the leading instability caused by
four-fermion couplings is the emergence of $\mathcal{T}$SB SC order.
Among the six candidates, the $\mathcal{T}$SB SC state with
$id_{xz}$-wave symmetry is the most favorable. It is also of
particular interest to generalize our method to identify the
leading instability of the analogous topological $\mathcal{T}$SB-driven transition
in the polar phase of $^3$He~\cite{Volovik2003Book}.

\section{Effective theory and RG analysis}

We consider 3D line-nodal noncentrosymmetric superconductors
characterized by the symmetry group $\mathcal{G} = C_{4v} \times
\mathcal{T} \times \mathcal{P}$. The generalization to other groups
is straightforward. The low-energy mean-field Hamiltonian of this
system is formally written as \cite{Matsuura2013NJP,Moon2017PRB}
\begin{eqnarray}
H_0 = \sum_{\mathbf{k}}\Psi^\dagger_{\mathbf{k}}
\left[h(\mathbf{k})\tau^z + \Delta(\mathbf{k})
\tau^x\right]\Psi_{\mathbf{k}},
\end{eqnarray}
where the four-component spinor is $\Psi^\dagger_{\mathbf{k}} =
(\psi^\dagger_{\mathbf{k}}, i\sigma^y\psi^T_{-\mathbf{k}})$ with
$\psi^\dagger_{\mathbf{k}} = (c^*_{\mathbf{k}\uparrow},
c^*_{\mathbf{k}\downarrow})$. Pauli matrices $\tau^{x,y,z}$ and
$\sigma^{x,y,z}$ act in the particle-hole space and spin space,
respectively. The $\tau^z$ term captures the fermionic degrees of
freedom in the non-SC phase \cite{Frigeri2004PRL,Brydon2011PRB,
Moon2017PRB}. The energy spectrum is given by
$h(\mathbf{k})=\epsilon(\mathbf{k})-\mu + \alpha
\mathbf{l}(\mathbf{k})\cdot \mathbf{\sigma}$, where the kinetic
energy $\epsilon(\mathbf{k})=-2t(\cos k_x + \cos k_y + \cos k_z)$.
Spin-orbital coupling is included by introducing the parameter
$\alpha$. The $\tau^x$-term corresponds to the SC pairing, and the
SC order parameter has the form \cite{Moon2017PRB}
$\Delta(\mathbf{k}) = \Delta_s + \Delta_t
\mathbf{d}(\mathbf{k})\cdot\sigma$, where $\Delta_s$ is an
even-parity spin-singlet gap and $\Delta_t$ an odd-parity
spin-triplet gap. The pairing term is chosen to be along the same
direction as that of spin-orbital coupling, i.e.,
$\mathbf{l}(\mathbf{k}) = \mathbf{d}(\mathbf{k})$. The amplitudes
$\Delta_s$ and $\Delta_t$ are taken to be real and positive due to
$\mathcal{T}$-symmetry \cite{Moon2017PRB}.

As aforementioned, $\mathcal{T}$SB alters the topology of the SC
state. $\mathcal{T}$SB can be realized by adding
an extra $\tau^y$ term. In Ref.~\cite{Moon2017PRB}, Han \emph{et
al.} considered one specific four-fermion coupling
$\left(\Psi^\dagger \tau^y \Psi\right)^2$ and made a mean-field
analysis to generate $\tau^y$ term. This coupling is certainly not
the only possibility, and other four-fermion couplings could be
present. To make an unbiased judgement, we consider the following
four couplings
\begin{eqnarray}
H_{\mathrm{int}} = \sum^3_{i=0}\int d^3\mathbf{x}u_i
\left(\Psi^\dagger \gamma^{i}\Psi\right)^2,\label{Eq_H_int}
\end{eqnarray}
where $u_{0,1,2,3}$ are coupling constants, $\gamma^0 \equiv \tau^0
= I_{2\times2}$, and $\gamma^{1,2,3} \equiv \tau^{x,y,z}$.
Mean-field treatment ignores the quantum fluctuations and cannot
determine the relative importance of the above four possible
couplings. Here we go beyond the mean-field level, and take into
account the quantum fluctuations by using the RG techniques. It will
become clear that, RG not only selects out the most important
coupling, but also pins down the leading ordering instability.

One can verify \cite{Matsuura2013NJP,Moon2017PRB} that the above SC
gap has a line-node structure. Gapless fermions are excited around
the nodal lines. To obtain the low-energy continuum limit, we follow
Ref.~\cite{Moon2017PRB} and implement the approximations
$\epsilon(\mathbf{k}) \rightarrow \frac{\mathbf{k}^2}{2m}$ with
$m=\frac{1}{2t}$ and
$\vec{l}(\mathbf{k})\approx(k_x,k_y,0)=\mathbf{k}_\perp$. It
suffices to focus on the upper nodal-ring, since the lower
nodal-ring is equivalent. Gathering the effective free Hamiltonian
$H_0\sim\Psi^\dagger_{\mathbf{k}} \left(v_z\delta
k_z\tau^z + v_p\delta k_\perp\tau^x\right)\Psi_{\mathbf{k}}$
and interactions (\ref{Eq_H_int}) gives rise to the total low-energy effective action
\begin{eqnarray}
S_{\mathrm{eff}} \!\!\!&=&\!\!\!\!\! \int_{\mathbf{k},\omega}
\Psi^\dagger_{\mathbf{k},\omega}(-i\omega+v_z\delta k_z\tau^z +
v_p\delta k_\perp\tau^x)\Psi_{\mathbf{k},\omega} \nonumber \\
&&\!\!\!\!\! +\sum^3_{i=0}u_i\prod^4_{j=1}\int_{\mathbf{k}_j,\omega_j}
\Psi^\dagger_{\mathbf{k}_1,\omega_1} \gamma^i
\Psi_{\mathbf{k}_2,\omega_2} \Psi^\dagger_{\mathbf{k}_3,\omega_3}
\gamma^i \Psi_{\mathbf{k}_4,\omega_4} \nonumber \\
&&\!\!\!\!\!\times\delta^{(3)}(\mathbf{k}_1+\mathbf{k}_2-\mathbf{k}_3 -
\mathbf{k}_4) \delta(\omega_1+\omega_2-\omega_3-\omega_4). \label{Eq_S_eff}
\end{eqnarray}
The free fermion propagator is given by $G_0^{-1}(k) =
-i\omega+v_z\delta k_z \tau^z + v_p\delta k_\perp \tau^x$.
Throughout the following calculations, we focus on the limit of
$|k|\ll k_F$ (with the size of nodal ring $k_F$) and as such adopt
the approximation for the form factor
$\mathcal{F}(\mathbf{k})\rightarrow \mathcal{F}(\theta_\mathbf{k})$.
This allows us to adopt the approximations $\sin k_x\approx
\cos\theta_{\mathbf{k}}$, $\sin k_y\approx \sin\theta_{\mathbf{k}}$,
$\cos k_x-\cos k_y\approx\cos2\theta_{\mathbf{k}}$, and $\sin
k_z\approx u^z$. We use $v_z$ to denote the fermion velocity along
$z$-axis, and $v_p$ the one within the $x$-$y$ plane. In addition,
we now can invoke the following approximation \cite{Moon2017PRB}
\begin{eqnarray}
\int_{\mathbf{k},\omega}\equiv \int\frac{d^3kd\omega}{(2\pi)^4}
\approx\int\frac{d\delta k_z}{2\pi} \int k_F\frac{d\delta
k_\perp}{2\pi}\int \frac{d\theta_\mathbf{k}}{2\pi}
\int\frac{d\omega}{2\pi},
\end{eqnarray}
with $\delta k^2_\perp = \delta k^2_x + \delta k^2_y$.

The RG flow equations of fermion velocities and four-fermion
interaction parameters are \cite{Shankar1994RMP,Huh2008PRB,
She2010PRB,Wang2011PRB-imp}
\begin{eqnarray}
\frac{dv_z}{dl} &=& \frac{1}{4\pi
v_z v_p} v_z (u_0+u_3),\label{Eq_RG_v_z}\\
\frac{dv_p}{dl} &=& \frac{1}{4\pi v_z
v_p}v_p (u_0+u_1), \label{Eq_RG_v_perp} \\
\frac{du_0}{dl} &=&\frac{8}{16\pi v_zv_p}
[u^2_0+u_{2}(\mathcal{E}u_{3}+\mathcal{F}u_{1})], \label{Eq_RG_u0}\\
\frac{du_1}{dl} &=&\frac{1}{16\pi v_zv_p}[u_{1}(1+
\mathcal{F}-\mathcal{E})(u_{0}+3u_{1}-u_{2}-u_{3})\nonumber\\
&&+8u_1(u_0+\mathcal{F}u_{3})-4u_{2}u_{3}],
\label{Eq_RG_u1} \\
\frac{du_2}{dl} &=&\frac{1}{16\pi v_zv_p}
[u_{2}(1+\mathcal{F}+\mathcal{E})(u_{0}-u_{1}+3u_{2}-u_{3})\nonumber\\
&& +8u_{2}(u_0+\mathcal{E}u_{1} + \mathcal{F}u_{3})-4u_{1}u_{3}],
\label{Eq_RG_u2} \\
\frac{du_3}{dl} &=&\frac{1}{16\pi v_z v_p}[u_{3}(1-\mathcal{F}
+\mathcal{E})(u_{0}-u_{1}-u_{2}+3u_{3}) \nonumber \\
&&+8u_3(u_0+\mathcal{E}u_{1})-4u_{1}u_{2}]. \label{Eq_RG_u3}
\end{eqnarray}
Here, the coefficients $\mathcal{E}$ and $\mathcal{F}$ as well as
the one-loop calculations of vertex corrections are presented in
Appendix~\ref{Appendix_u_i}.

\begin{figure}
\centering
\includegraphics[width=4.8in]{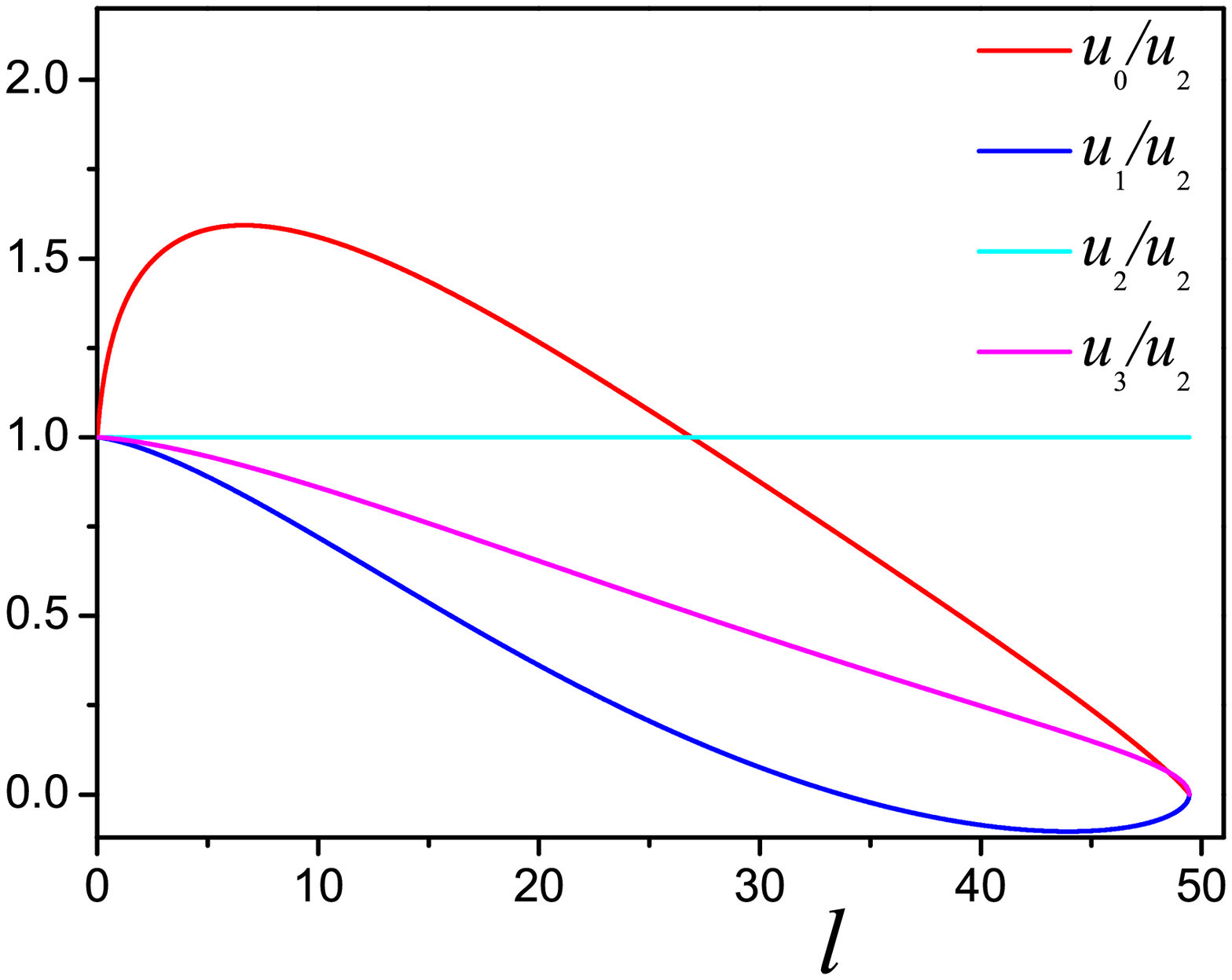}\\
\vspace{-7.9cm}
\hspace{0.5cm}\includegraphics[width=1.9in]{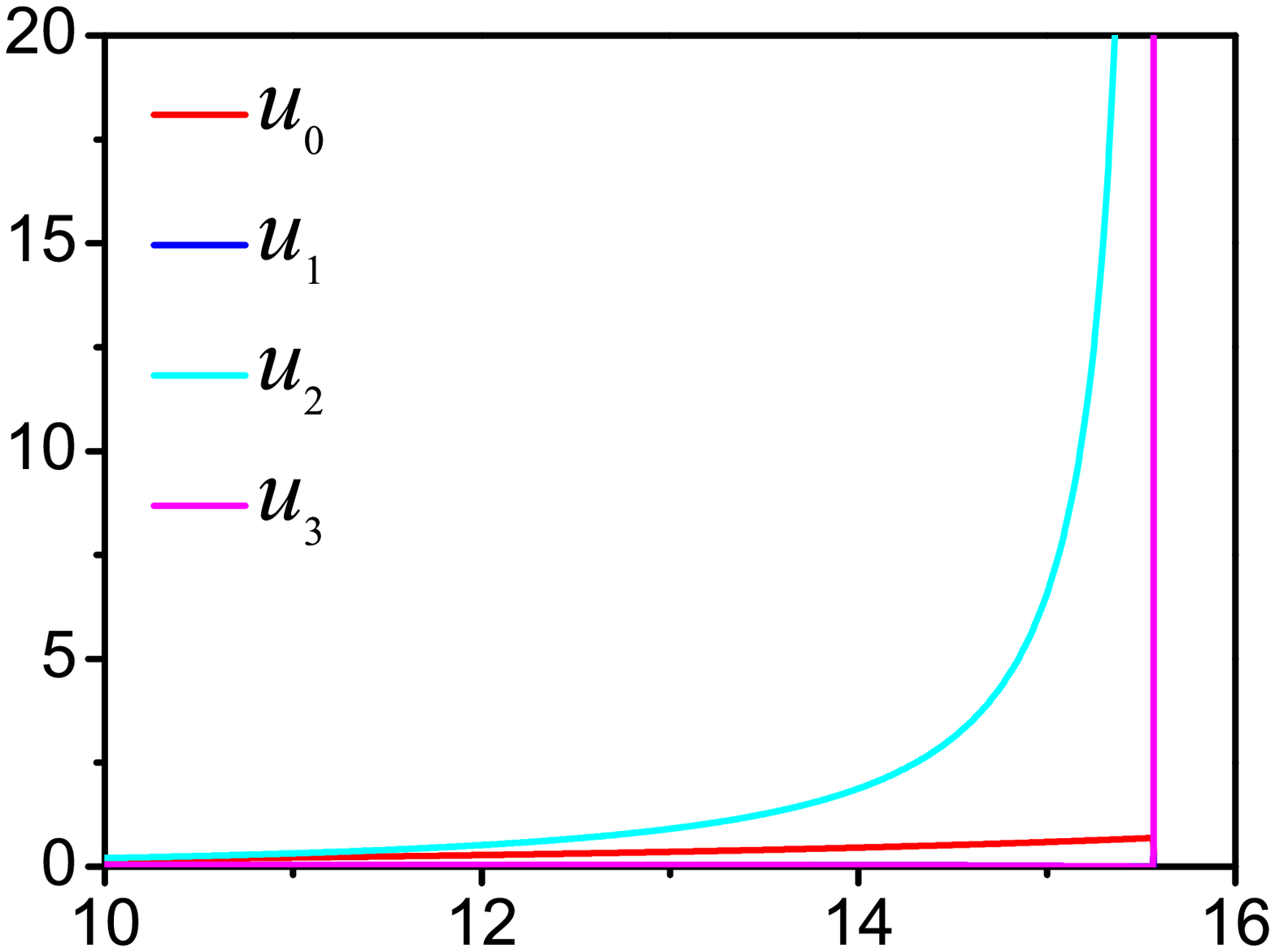}\\
\vspace{2.3cm} \caption{(Color online) Evolutions of $u_i/u_2$
($i=0,1,2,3$) with $\frac{v_z(0)}{v_p(0)}=5$ and $u_i(0)=10^{-4}$
and $u_i$ (inset) with $u_i(0)=10^{-4}$ and $v_z(0)=v_p(0)=0.001$.
This fixed point is independent of the initial values of fermion
velocities and coupling constants (the curves for $u_1$ and $u_3$
coincide).} \label{Fig_Relative-FP}
\end{figure}

\section{Results and discussions}

After solving the flow equations (\ref{Eq_RG_v_z})-(\ref{Eq_RG_u3}),
we find that all of the four parameters $u_i$ diverge at certain
critical scale $l_c$, no matter whether the starting values of $v_z$
and $v_p$ are equal or not. The results are shown in the inset of
Fig.~\ref{Fig_Relative-FP}. The four-fermion couplings are thus
marginally relevant at low energies \cite{Shankar1994RMP} and can
lead to some kinds of phase-transition instability
\cite{Vafek2014PRB,Wang2017QBCP,Wang2018JPCM,
Chubukov2016PRX,Wang2014PRB}.

The parameters $u_i$ go to infinity at different speeds. Their
relative importance \cite{Vafek2012PRB,Vafek2014PRB,Wang2017QBCP}
can be distinguished by re-scaling all the parameters with respect
to one of them (no sign change). Here, we choose to analyze the
energy-dependence of the ratios $u_i/u_2$. At the outset, we let
$v_z$ be equal to $v_p$ at $l = 0$. Numerical analysis reveals that
the re-scaled parameters $u_i/u_2$ inescapably flow to
$(u_0,u_1,u_2,u_3)/u_2 \approx(0,0,1,0)$. This result is independent
of the bare values of $u_i(0)$, $v_z(0)$, and $v_p(0)$, as
illustrated in Fig.~\ref{Fig_Relative-FP}. We then consider the case
in which the bare fermion velocities are not equal, i.e.,
$v_z(0)\neq v_p(0)$. As can be seen from Fig.~\ref{Fig_Relative-FP},
the relative fixed point of $u_i/u_2$ is considerably robust against
the velocity anisotropy. In particular, the fixed point
$(u_0,u_1,u_2,u_3)/u_2 \approx(0,0,1,0)$ is still present. Clearly,
the coupling $\left(\Psi^\dagger\tau^y \Psi\right)^2$ is more
important at low energies than the rest three ones.

According to traditional notion, the runaway flow of
$\left(\Psi^\dagger\gamma^i \Psi\right)^2$ implies that the system
becomes unstable and the fermion bilinear $\Psi^\dagger\gamma^i
\Psi$ acquires a finite vacuum expectation value. However, it has
recently becoming clear that this connection might not be always
correct \cite{Metzner2000PRL,Vafek2012PRB,Chubukov2016PRX}. Even
if the coupling $\left(\Psi^\dagger\tau^y \Psi\right)^2$ dominates
in the low-energy region, the transition driven by nonzero $\langle
\Psi^\dagger\tau^y \Psi\rangle$ is not necessarily the leading
instability. Other instabilities may be more favorable. Moreover,
the nonzero $\langle \Psi^\dagger\tau^y \Psi\rangle$ corresponds to
several possible states. One powerful strategy is to consider all of
the instabilities at the same time and calculate the energy
dependence of all the corresponding susceptibilities
\cite{Metzner2000PRL,Vafek2012PRB,Chubukov2016PRX}. The leading
instability should be the one whose susceptibility diverges most
rapidly as energy is lowered \cite{Metzner2000PRL,Vafek2012PRB,
Chubukov2016PRX}. It is therefore necessary to take into account all
the four types of four-fermion couplings defined in
Eq.~(\ref{Eq_H_int}).

The mean value $\langle \Psi^\dagger\tau^0 \Psi\rangle$ must vanish
owing to the particle-hole symmetry. The nonzero mean value $\langle
\Psi^\dagger\tau^z \Psi\rangle$ leads to a shift in the
quasiparticle energy. A nonzero $\langle \Psi^\dagger\tau^x
\Psi\rangle$ generates a constant shift to the spin-singlet gap
$\Delta_s$, which then changes the radius of the gap-vanishing nodal
line, but preserves its circular shape. The nonzero mean value
$\langle \Psi^\dagger\tau^y \Psi\rangle$ dynamically breaks the
$\mathcal{T}$-symmetry, and as such alters the topology.

After performing calculations, we have verified that the generation
of nonzero $\langle \Psi^\dagger\tau^y \Psi\rangle$ is more
favorable than the rest ones. However, symmetry consideration allows
for six candidate $\mathcal{T}$SB SC states \cite{Moon2017PRB}, as
shown in Table~\ref{table_source}. While in principle each of the
six states could occur, only one of them could be realized at low
energies. To identify the leading instability, we need to evaluate
the susceptibility \cite{Vafek2012PRB,Vafek2014PRB,Wang2017QBCP}
associated with each possible order as the system approaches the
relative fixed point $(u_0,u_1,u_2,u_3)/u_2 \approx(0,0,1,0)$. For
this purpose, we introduce a source term \cite{Chubukov2010PRB,
Vafek2014PRB,Chubukov2016PRX}, and write down the following
source-related effective action
\begin{eqnarray}
S_{\mathrm{source}} &=& \int d\tau\int d^3\mathbf{x}
\left(\sum^6_{i=1}\Delta_i\Psi^\dagger \mathcal{M}_i
\Psi\right).\label{Eq_S_source}
\end{eqnarray}
Here, the concentration is on the charge-channel since the six
instabilities are all closely related to the dynamically generated
mass term. Each $\Delta_i$ represents an SC order parameter, and
matrix $\mathcal{M}_i$ specifies how $\Delta_i$ couples to the
fermionic degrees of freedom. The connection between $\Delta_i$ and
$\mathcal{M}_i$ is summarized in Table~\ref{table_source}. Combining
the effective action (\ref{Eq_S_eff}) and the additional source term
(\ref{Eq_S_source}) yields a renormalized effective action
\begin{eqnarray}
S'_{\mathrm{eff}} = S_{\mathrm{eff}} + S_{\mathrm{source}}.
\label{Eq_S_eff_new}
\end{eqnarray}

\begin{figure}
\centering
\includegraphics[width=3.7in]{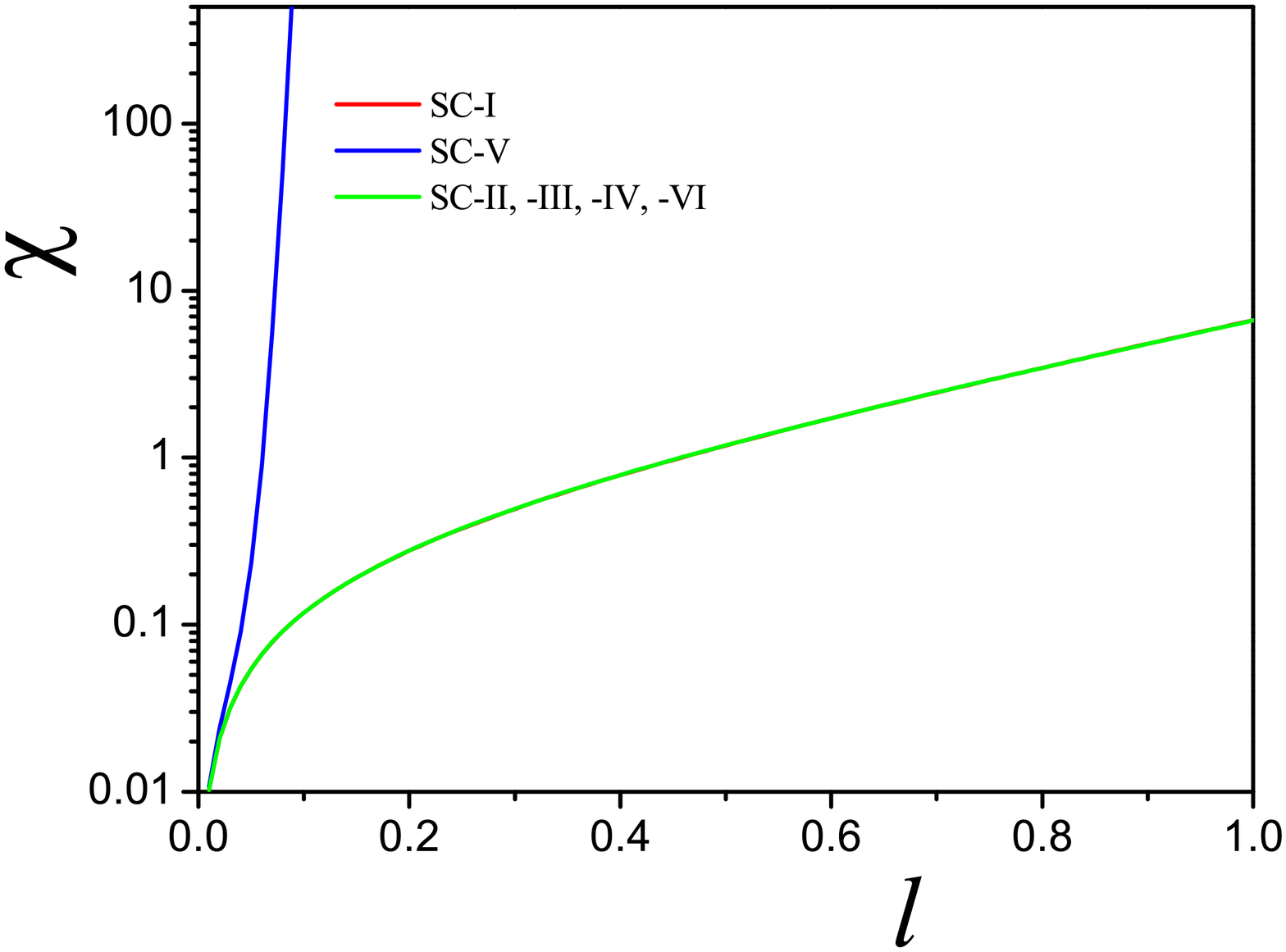}\\
\vspace{-3.5cm}
\hspace{3.6cm}\includegraphics[width=1.39in]{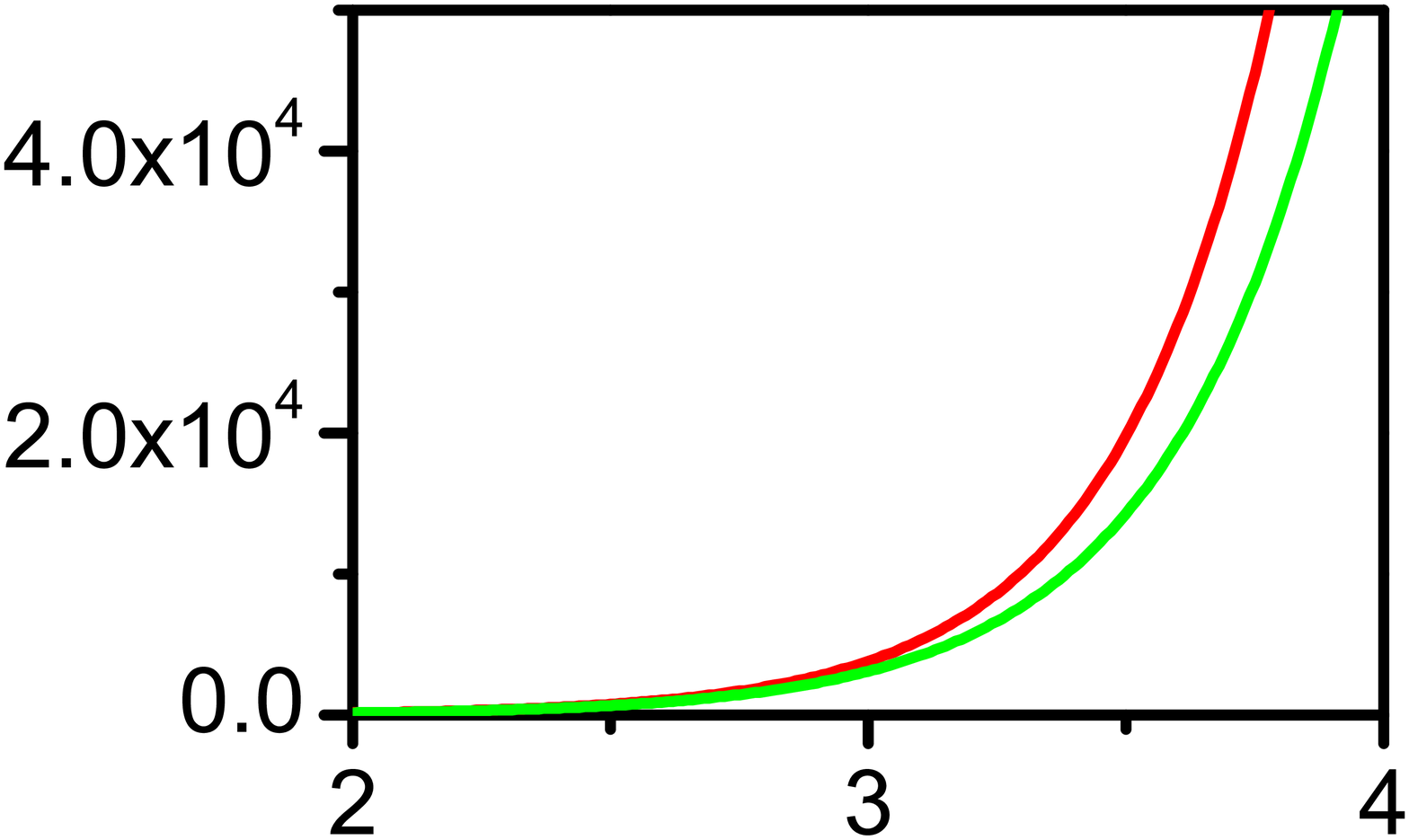} \\
\vspace{0.90cm} \caption{(Color online) Evolutions of the
susceptibilities of all the possible instabilities. We choose
$u_i(0) = 10^{-4}$ for $i=0,1,2,3$ and $v_z(0)=v_p(0)=0.02$. Inset:
the splitting of susceptibilities that are nearly indistinguishable
nearby the leading instability. The basic results are independent of
the initial values of fermion velocities and fermion-fermion
interactions. The details of these orders are given in
Table~\ref{table_source}.} \label{chi_u-1M4_aniso}
\end{figure}

On the basis of Eq.~(\ref{Eq_S_eff_new}), one can directly compute
the one-loop corrections to the source strength. We find that
$\Delta_i$ depend on $l$ as follows
\begin{eqnarray}
\frac{d\ln\Delta_i}{dl} =
\mathcal{F}_i(v_z,v_p,u_0,u_1,u_2,u_3,\Delta_i),
\label{Eq_RG_source}
\end{eqnarray}
where the concrete expressions of $\mathcal{F}_i$ are given in
Appendix~\ref{Sec_RG_analysis}. Susceptibilities are computed via the
following formula \cite{Nelson1975PRB,Vafek2012PRB,Vafek2014PRB,
Wang2017QBCP}
\begin{eqnarray}
\delta\chi_i(l) = -\frac{\partial^2 \delta f}{\partial \Delta_i(0)
\partial\Delta_i^*(0)},\label{Eq_chi}
\end{eqnarray}
where $\chi_i$ denotes the susceptibility and
$f(\Delta_i,\Delta_i^*)$ is the free energy functional. The
susceptibilities are calculated based on the RG flows of interaction
parameters (\ref{Eq_RG_u0})-(\ref{Eq_RG_u3}) and fermion velocities
(\ref{Eq_RG_v_z})-(\ref{Eq_RG_v_perp}), as well as the strength of
source terms (\ref{Eq_RG_source}).

Numerical computations show that the leading instability is the
transition into $\mathcal{T}$SB SC-V state, whose gap symmetry is
$id_{xz}$-wave. The results are displayed in
Fig.~\ref{chi_u-1M4_aniso}. When the scale $l$ grows, the
susceptibility of SC-V state increases rapidly and diverges at
certain finite value $l_c$. As illustrated in the inset of
Fig.~\ref{chi_u-1M4_aniso}, all the other susceptibilities are still
relatively small at $l_c$. Notice that the six $\mathcal{T}$SB SC
states are incompatible: only one can exist at low energies. Once
SC-V state is realized, the rest five SC states are excluded. We
have verified that this conclusion is insensitive to the starting
values of fermion velocities and four-fermion coupling parameters.

A proper extension of our approach may be also applicable
to the polar phase of $^3$He. Given the polar phase and phase-A of $^3$He share
analogous topological structures of Fermi surfaces with their Phase-I
and Phase-II counterparts sketched in Fig.~\ref{Fig_phase}~\cite{Volovik2003Book,Moon2017PRB},
one can realize that principal effective low-energy action of the polar phase of $^3$He
is similar to that of 3D line-nodal noncentrosymmetric superconductor.
It is therefore natural to expect that the interactions among these gapless
quasiparticles excited from nodal points/lines would play a crucial role
in pinning down the concrete states in the polar phases
once the system undergoes some phase transition from
the phase-A to polar phase of $^3$He.

\section{Summary}

We have studied the impact of short-range four-fermion
interactions on the low-energy properties of three-dimensional
line-nodal superconductors. In the absence of such short-range
interactions, the SC state is topological and has line nodes, which
is protected by $\mathcal{T}$ symmetry. We have found that
short-range interactions significantly alter the dynamics of the
gapless nodal fermions. Additionally, we have showed that,
short-range interactions can lead to dynamical breaking of
$\mathcal{T}$ symmetry, which then changes the topology of the SC
state and turns the system into a distinct $\mathcal{T}$SB SC phase.
After calculating the susceptibilities associated with six candidate
$\mathcal{T}$SB SC phases, we have demonstrated that the SC phase
with $\cos(\theta_{\mathbf{k}})\tau^y $ coupling matrix, which
mounts to the SC state having $id_{xz}$-wave pairing in the
noncentrosymmetric case and $p_z+ip_x$-wave pairing in the
$p_z$-polarized case, is the leading instability. We expect that
future experiments probe the existence of such a $\mathcal{T}$SB SC
state in unconventional line-nodal superconductors.

\section*{ACKNOWLEDGEMENTS}

I am very grateful to Prof. G. -Z. Liu for providing
invaluable comments and kindly polishing this manuscript.
In addition, J.W. acknowledges Dr. D.- V. Efremov, and Dr. C. Ortix, as well as
Prof. J. van den Brink for correlated collaborations and helpful correspondence.
The author would also like to thank Prof. X.-Y. Pan and Miss Y. -H. Zhai for useful
discussions. This work is partially supported by the National Natural
Science Foundation of China under Grant No. 11504360.

\appendix

\section{One-loop corrections to interaction parameters}
\label{Appendix_u_i}

Here, we provide details for the one-loop RG calculations.
As delineated in Fig.~\ref{Fig_fermion_propagator_correction} and
Fig.~\ref{Fig_fermion_interaction_correction}, the four-fermion
interaction contributes to both the fermion self-energy and the
four-fermion vertices. The one-loop fermion self-energy is given by
Eq.~(\ref{Eq_Sigma}). Hereafter, we focus on the one-loop
renormalization of four-fermion vertices. After evaluating the
Feynman diagrams shown in
Fig.~\ref{Fig_fermion_interaction_correction}, we derive the
corrections to vertices as follows,
\begin{widetext}
\begin{eqnarray}
\delta S^{i-v}_{u_0} &=& \int_{\mathbf{k}_1,\omega_1}
\int_{\mathbf{k}_2,\omega_2}\int_{\mathbf{k}_3,\omega_3}
\psi^\dagger(\omega_1,\mathbf{k}_1)\tau^0\psi(\omega_2,\mathbf{k}_2)
\psi^\dagger(\omega_3,\mathbf{k}_3) \tau^0\psi(\omega_1 +
\omega_2-\omega_3,\mathbf{k}_1+\mathbf{k}_2-\mathbf{k}_3)\nonumber
\\
&&\times\left[\frac{4u_{2}(u_{3}\mathcal{E}+u_{1}
\mathcal{F})}{8\pi v_zv_p}l\right],\\
\delta S^{i-v}_{u_1} &=& \int_{\mathbf{k}_1,\omega_1}
\int_{\mathbf{k}_2,\omega_2}\int_{\mathbf{k}_3,\omega_3}
\psi^\dagger(\omega_1,\mathbf{k}_1)\tau^1\psi(\omega_2,\mathbf{k}_2)
\psi^\dagger(\omega_3,\mathbf{k}_3) \tau^1\psi(\omega_1 +
\omega_2-\omega_3,\mathbf{k}_1+\mathbf{k}_2-\mathbf{k}_3)\nonumber
\\
&&\times\left[\frac{u_{1}(u_{0}+3u_{1}-u_{2}-u_{3})(\frac{1}{2}+
\mathcal{F}-\mathcal{E})-2u_{2}u_{3}+4u_{1}u_{3}\mathcal{F}}{8\pi v_zv_p}l\right],\\
\delta S^{i-v}_{u_2} &=& \int_{\mathbf{k}_1,\omega_1}
\int_{\mathbf{k}_2,\omega_2}\int_{\mathbf{k}_3,\omega_3}
\psi^\dagger(\omega_1,\mathbf{k}_1)\tau^2\psi(\omega_2,\mathbf{k}_2)
\psi^\dagger(\omega_3,\mathbf{k}_3) \tau^2\psi(\omega_1 +
\omega_2-\omega_3,\mathbf{k}_1+\mathbf{k}_2-\mathbf{k}_3)\nonumber
\\
&&\times\left[\frac{u_{2}(u_{0}-u_{1}+3u_{2}-u_{3})(\frac{1}{2}+\mathcal{E}
+\mathcal{F})-2u_{1}u_{3}
+4u_{2}(u_{3}\mathcal{F}+u_{1}\mathcal{E})}{8\pi v_zv_p}l\right],\\
\delta S^{i-v}_{u_3} &=& \int_{\mathbf{k}_1,\omega_1}
\int_{\mathbf{k}_2,\omega_2}\int_{\mathbf{k}_3,\omega_3}
\psi^\dagger(\omega_1,\mathbf{k}_1)\tau^3\psi(\omega_2,\mathbf{k}_2)
\psi^\dagger(\omega_3,\mathbf{k}_3) \tau^3\psi(\omega_1+\omega_2 -
\omega_3,\mathbf{k}_1+\mathbf{k}_2-\mathbf{k}_3)\nonumber \\
&&\times\left[\frac{u_{3}(u_{0}-u_{1}-u_{2}+3u_{3})(\frac{1}{2}-\mathcal{F}
+\mathcal{E})-2u_{1}u_{2}
+4u_{3}u_{1}\mathcal{E}} {8\pi v_zv_p}l\right],
\end{eqnarray}
with $\mathcal{E}$ and $\mathcal{F}$ being designated as
\begin{eqnarray}
\mathcal{E}&\equiv&\frac{1}{\pi}\int_{-\frac{\pi}{2}}^{\frac{\pi}{2}}d\theta
\frac{v_{z}v^{3}_{\bot}\cos^{2}\theta}
{(v_{z}^{2}\sin^{2}\theta+v_{\bot}^{2}\cos^{2}\theta)^
{\frac{3}{2}}},\hspace{0.3cm}
\mathcal{F}\equiv\frac{1}{\pi}\int_{-\frac{\pi}{2}}^{\frac{\pi}{2}}
d\theta\frac{v^{3}_{z}v_{\bot}\sin^{2}\theta}
{(v_{z}^{2}\sin^{2}\theta+v_{\bot}^{2}\cos^{2}\theta)^
{\frac{3}{2}}}.
\end{eqnarray}

\begin{figure}
\centering
\includegraphics[width=5.0in]{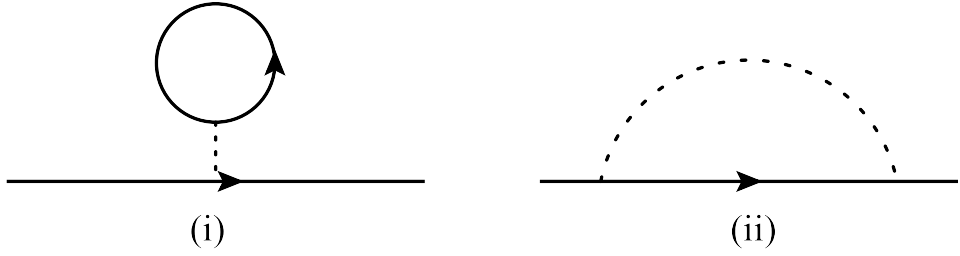}
\vspace{-0.1cm} \caption{One-loop corrections to the fermion
propagator (the dashed line indicates the
interactions).}\label{Fig_fermion_propagator_correction}
\end{figure}

\begin{figure}
\centering
\includegraphics[width=5.0in]{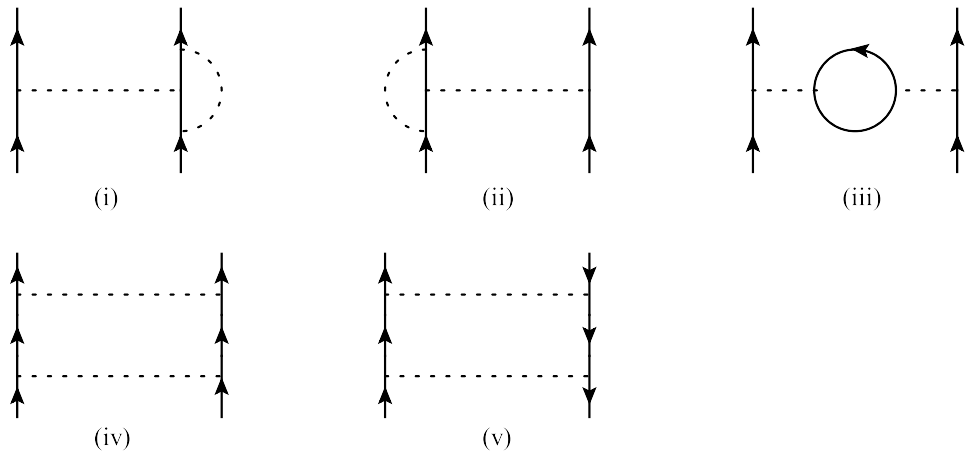}
\vspace{-0.1cm} \caption{One-loop corrections to the fermion
interacting couplings (the dashed line indicates the
interactions).}\label{Fig_fermion_interaction_correction}
\end{figure}

\end{widetext}

\begin{figure}
\centering
\includegraphics[width=2.6in]{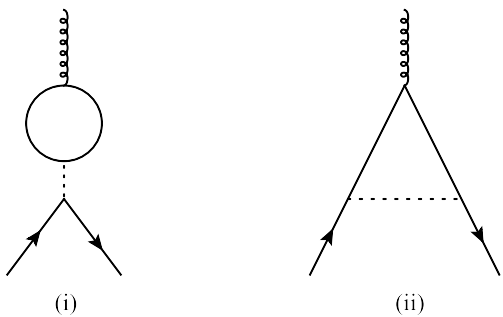}
\vspace{-0.1cm} \caption{One-loop corrections to the strength of
source terms (the dashed line indicates the
interactions).}\label{Fig_source_correction}
\end{figure}

\section{Renormalization group analysis}\label{Sec_RG_analysis}

Within the standard formalism of Wilsonian RG approach
\cite{Wilson1975RMP,Polchinski9210046,Shankar1994RMP}, we need to
integrate out the fields defined in the momentum shell $b\Lambda < k
< \Lambda$, where $b<1$, to determine how the interaction parameters
of the effective theory~(\ref{Eq_S_eff}) evolve upon
lowering the energy scale. To proceed, we introduce $\Lambda$ as the
upper energy scale and the variable parameter $b=e^{-l}$, where
$l>0$ is a varying parameter \cite{Wilson1975RMP,Polchinski9210046,
Shankar1994RMP}. One can re-scale momenta and energy by $\Lambda_0$,
which is determined by the inverse lattice constant, i.e.
$k\rightarrow k/\Lambda_0$ and $\omega\rightarrow
\omega/\Lambda_0$~\cite{Huh2008PRB,She2010PRB,Wang2011PRB-imp}. This
manipulation will help simplify RG calculations.

Before going further, we need to first specify the RG re-scaling
transformations. In light of the spirit of momentum-shell RG theory,
the $(-i\omega)$ term in Eq.~(3) in the main text can be identified
as the free fixed point. Guided by this principle, it is now easy to
verify that the momenta, energy and fermion fields should transform
as follows \cite{Shankar1994RMP,Huh2008PRB,She2010PRB,
Wang2011PRB-imp}
\begin{eqnarray}
\omega' &=& \omega e^{-l},\\
\delta k'_z &=& \delta k_ze^{-l},\\
\delta k'_\perp &=& \delta k_\perp e^{-l},\\
\Psi'_{\mathbf{k'},\omega'} &=& \Psi_{\mathbf{k},\omega}
e^{\frac{1}{2}\int^l_0 dl(4-\eta_f)}.
\end{eqnarray}
Here, a parameter $\eta_f$ is implemented to represent the fermion
anomalous dimension, which would be generated by the four-fermion
interactions.

Making use of the above definitions, we can perform RG calculation
up to the one-loop order. As delineated in Appendix~\ref{Appendix_u_i}
the four-fermion interactions contribute to both the fermion
self-energy and the four-fermion vertex functions. One-loop
correction to fermion self-energy is
\begin{widetext}
\begin{eqnarray}
\Sigma(\omega,\mathbf{k})  &=&
\int_{\mathbf{k},\omega}\Psi^\dagger(\omega,\mathbf{k})
\Bigl[-(u_0\mathcal{C}l)i\omega\tau_0 - (u_1\mathcal{C}l)v_p\delta
k_\perp\tau^x -(u_3\mathcal{C}l)v_z\delta k_z\tau^z \Bigr]
\Psi(\omega,\mathbf{k}), \label{Eq_Sigma}
\end{eqnarray}
\end{widetext}
where parameter $\mathcal{C}$ is given by
\begin{eqnarray}
\mathcal{C} = \frac{1}{4\pi v_zv_p}.
\label{Eq_coeff_C}
\end{eqnarray}
Gathering the above self-energy and keeping
$\Psi^\dagger(-i\omega)\Psi$ invariant under RG transformations, we
get
\begin{eqnarray}
\eta_f = -\frac{u_0}{4\pi v_zv_p}.\label{Eq_eta_f}
\end{eqnarray}


The one-loop contributions to the fermion-source coupling are plotted
in Fig.~\ref{Fig_source_correction}. Repeating similar RG procedures,
we find that the strength parameters of source
terms satisfy the following coupling equations
\begin{widetext}
\begin{eqnarray}
\frac{d\ln\Delta_\mathrm{I}}{dl}
&=&\left[(1-\eta_f)+\frac{2(u_0-u_2)(2\mathcal{E}+2\mathcal{F}+1)+
2(u_1+u_3)(2\mathcal{E}+2\mathcal{F}-1)}{32\pi v_zv_\perp}\right]\Delta_\mathrm{I},\\
\frac{d\ln\Delta_\mathrm{II}}{dl}
&=&(1-\eta_f)\Delta_\mathrm{II},\\
\frac{d\ln\Delta_\mathrm{III}}{dl}&=&(1-\eta_f)\Delta_\mathrm{III},\\
\frac{d\ln\Delta_\mathrm{IV}}{dl}&=&(1-\eta_f)\Delta_\mathrm{IV},\\
\frac{d\ln\Delta_\mathrm{V}}{dl}
&=&\left[(1-\eta_f)+\frac{4(u_0-u_2)(\mathcal{E'}+\mathcal{F'}+\mathcal{C'})+
4(u_1+u_3)(\mathcal{E'}+\mathcal{F'}-\mathcal{C'})}{32\pi v_zv_\perp}\right]\Delta_\mathrm{V},\\
\frac{d\ln\Delta_\mathrm{VI}}{dl}&=&(1-\eta_f)\Delta_\mathrm{VI},
\end{eqnarray}
\end{widetext}
with $\eta_f$ being denominated in Eq.~(\ref{Eq_eta_f}) and $\mathcal{C'}$, $\mathcal{E'}$,
$\mathcal{F'}$ introduced as
\begin{eqnarray}
\mathcal{C'}&=&\frac{1}{\pi} \int_{-\frac{\pi}{2}}^{\frac{\pi}{2}}d\theta\frac{v_{z}v_{\bot}\cos\theta }{\sqrt{v^{2}_{z}\sin^{2}\theta+v^{2}_{\bot}\cos
^{2}\theta}},\\
\mathcal{E'}&=&\frac{1}{\pi}\int_{-\frac{\pi}{2}}^{\frac{\pi}{2}}d\theta
\frac{v_{z}v^{3}_{\bot}\cos^{3}\theta}
{(v_{z}^{2}\sin^{2}\theta+v_{\bot}^{2}\cos^{2}\theta)^
{\frac{3}{2}}},\\
\mathcal{F'}
&=&\frac{1}{\pi}\int_{-\frac{\pi}{2}}^{\frac{\pi}{2}}
d\theta\frac{v^{3}_{z}v_{\bot}\cos\theta\sin^{2}\theta}
{(v_{z}^{2}\sin^{2}\theta+v_{\bot}^{2}\cos^{2}\theta)^
{\frac{3}{2}}}.
\end{eqnarray}
Here, the strength of source terms $\Delta_i$ with $i=1-9$ is presented in
Table~\ref{table_source} of the main text.


\end{document}